# Self-assembled nematic colloidal motors powered by light

Ye Yuan[1], Ghaneema N. Abuhaimed[1], Qingkun Liu[1] and Ivan I. Smalyukh[1,2,3*]

[1]*Department of Physics, University of Colorado, Boulder, CO 80309, USA*
[2]*Soft Materials Research Center and Materials Science and Engineering Program, Department of Electrical, Computer, and Energy Engineering, University of Colorado, Boulder, CO 80309, USA*
[3]*Renewable and Sustainable Energy Institute, National Renewable Energy Laboratory and University of Colorado, Boulder, CO 80309, USA*
*\*Correspondence to: ivan.smalyukh@colorado.edu*

**Abstract.**

**Biological motors are marvels of nature that inspire creation of their synthetic counterparts with comparable nanoscale dimensions, high efficiency and diverse functions. Molecular motors have been synthesized, but obtaining nanomotors though self-assembly remains challenging. Here we describe a self-assembled colloidal motor with a repetitive light-driven rotation of transparent micro-particles immersed in a liquid crystal and powered by a continuous exposure to unstructured ~1 nW light. A monolayer of azobenzene molecules defines how the liquid crystal's optical axis mechanically couples to the particle's surface, as well as how they jointly rotate as the light's polarization changes. The rotating particle twists the liquid crystal, which changes polarization of traversing light. The resulting feedback mechanism yields a continuous opto-mechanical cycle and drives the unidirectional particle spinning, with handedness and frequency robustly controlled by polarization and intensity of light. Our findings may lead to opto-mechanical devices and colloidal machines compatible with liquid crystal display technology.**

**Introduction.**

Miniaturization of motors and machines down to nanometer and micrometer scales could literally lead to a new industrial revolution, but remains a challenge[1]. Man-made nanomotors were first envisaged by Feynman[1,2] and are key to many future applications[3-7]. Feringa chemically synthesized light-driven molecular motors[8], which were later used in synthetic nanocars[9] and in controlling chiral liquid crystals (LCs) with colloidal inclusions[10]. Much like in a four-stroke cycle combustion engine, the directional rotation of Feringa's motors stems from driving the molecule between a series of four conformations selected by cycling temperature and exposure to light[8,9]. Rotation of shape-anisotropic or birefringent particles can be also achieved by harnessing optical angular momentum transfer owing to the change in polarization of transmitted light[11], but this requires high-power laser traps. Many organic molecules, such as azobenzenes[12], exhibit facile polarization-sensitive responses to light, but their utility in the designs of artificial motors is limited[1]. Similarly, LCs are known for their fascinating ability of controlling polarization and intensity of light, with plethora of technological applications, such as the flat panel displays[13], but LC-enabled motors or machines capable of converting external sources of energy into mechanical work are rarely demonstrated[1,10].

Biological motors consume energy and convert it into motion or mechanical work, often with efficiency superior to that of the best combustion engines and electric motors[14]. This is achieved by many self-assembled biomolecules that cooperatively drive essential biological processes like flagellum rotation in bacteria, DNA packing in capsids of viruses and crawling of cells[14]. However, development of such self-assembled motors in purely synthetic systems is a grand challenge in colloidal science and nanotechnology.

To achieve such cooperative light-driven work, we let hundreds of millions of azobenzene molecules self-assemble into monolayers on surfaces of thin colloidal platelets, which are then dispersed in a nematic LC. This yields a colloidal light-driven motor with a highly controllable repetitive rotation of the micro-particle powered by low-intensity unstructured light (~1 nW). Self-assembled azobenzene molecules[12] mechanically couple the LC's optical axis to the particle's surface, mediating their joint rotation in response to changes of the light's polarization. Therefore, when the particle rotates, it causes LC twisting and further changes of light polarization[13], resulting in a feedback mechanism that leads to a continuous opto-mechanical cycle and unidirectional particle spinning. We characterize and discuss the physical underpinnings behind this fascinating behavior.

**Results**

**Structure and principles of operation.** Our colloidal motors are formed by thin, optically transparent silica platelets, with monolayers of azobenzene molecules self-assembled on their surfaces, which are immersed in a nematic LC (see Methods and Fig. 1). Differing from colloids in conventional isotropic fluids hosts[15], nematic colloidal platelets define surface boundary conditions for the surrounding molecules, spontaneously align with the large-area faces parallel to the sample plane and perturb the long-range orientational order of the LC described by a director **n**, the average local direction of rod-like molecular ordering[15,16]. These distortions manifest themselves through appearance of bright brushes around the platelet edges when viewed under a microscope between crossed polarizers (Fig. 1a), which correspond to the spatial regions where **n** departs from the uniform far-field director **n**$_0$ of an aligned LC. Locally discontinuous bending of director field **n(r)** around these particles results in formation of surface point defects, called "boojums", at the particle's poles along $n_0$ (Fig. 1a-c)[17]. Such equilibrium director structures around colloidal particles undergoing Brownian motion are stable when observed using red (or infrared) light, and when white or blue imaging or excitation light has linear polarization orthogonal to $n_0$ (Fig. 1a-c).

Remarkably, low-intensity ~1 nW blue excitation light with polarization states $P_e$ differing from the linear polarization perpendicular to $n_0$ drives many forms of particle dynamics, ranging from continuous rotation (Figs. 1d-g and 2 and Supplementary Movies 1-3) to periodic orientational oscillations (Fig. 3 and Supplementary Movies 4, 5). When the particle rotates, additional bright regions appear in the polarizing optical micrographs taken between crossed polarizers (Fig. 1d), indicating twist of $n(r)$ across the cell above and below the platelet (Fig. 1e-g). At the large-area faces of the particle, trans-state azobenzene moieties tend to align parallel to the faces and perpendicular to $P_p$ [12], with their orientation mechanically coupled to $n(r)$ through surface anchoring, where $P_p$ is the orientation of the long axis of polarization ellipse of light traversing the platelet. Excitation light with linear polarization $P_e \perp n_0$ traverses the sample in an ordinary mode without changing its polarization state[13], allowing for an equilibrium "stationary" configuration of a colloidal platelet having large-area faces parallel to $n_0$ and azobenzene moieties orthogonal to $P_p \parallel P_e$ (Fig. 1a-c). Light with all other polarization states $P_e$ causes a cooperative alignment and rotation of the azobenzene moieties within the monolayer molecules that remain anchored to platelet surfaces, which, in turn, rotate $n(r)$ and at the platelet-LC interfaces together with the platelet itself. To meet the boundary conditions on the rotating particle surfaces and the confining substrates, $n(r)$ twists across the cell (Fig. 1g) and modifies the polarization state of light traversing the platelet, thus prompting additional rotation of azobenzene moieties, platelet and $n(r)$ to yield an opto-mechanical cycle. This nonlinear feedback can be tuned to obtain desired dynamical responses of particles, ranging from pendulum-like orientational oscillations to a continuous unidirectional rotation (Figs. 1-3 and Supplementary Movies 1-5). Remarkably, this out-of-equilibrium behavior stems from the spontaneously emerging cooperative action of self-assembled colloidal platelet, ~$10^8$ azobenzene molecules and ~$10^{12}$ of LC molecules associated with distortions of **n(r)** both across the cell thickness and in lateral directions. Operation of our light-driven motor would be impossible without the LC host medium. In fact, the platelet rotation stops completely when LC is heated to the isotropic phase (>35 °C for the used LC host). Key to enabling the repetitive particle rotation are optical, surface anchoring and viscoelastic properties of the LC host medium. These properties allow for smooth and dynamic spatial deformation of the optical axis directly related to the dynamic polarization changes of light traversing the platelet[13], cooperative mutual alignment of LC and azobenzene dye molecules at surfaces that mechanically couple **n(r)** to the platelet through surface anchoring and orientational elasticity[12], as well as the strong response of azobenzene moieties to weak external stimuli[17]. Below we characterize and model details of this behavior.

**Characterization and physical underpinnings.** The high sensitivity of our motors to blue and violet light (Fig. 2) and lack of response to red light agree with the absorbance spectrum of trans-state azobenzene molecules (see Methods)[12], though ambient white light causes out-of-equilibrium colloidal dynamics too. Rotation of azobenzene moieties to be orthogonal to $P_p$ in the trans-state can be understood in simple terms as driven by their tendency to avoid the excited state[18]. Continuous particle rotation exhibits a threshold-like behavior that depends on polarization of incident light, with the onset of full rotation at typical optical powers of ~1 nW per particle (Fig. 2b), of which only ~0.5% is absorbed by self-assembled azobenzene monolayers (see Methods). The average angular speed of platelets shows a tendency to saturate with increasing optical power (Fig. 2b and Fig. 4). Dependent on the light intensity, the instantaneous angular speed exhibits rich periodic variations with periodicity often equal to about half that of particle rotation (Fig. 2a,c and d). Polarizing video microscopy reveals that this dynamics correlates with the behavior of **n(r)** coupled to the easy axis set by the average orientation of azobenzene moieties at the LC interface with a rotating platelet (Fig. 1d-f). In addition to this nonlinear dynamics of **n(r)** at the large-area faces, the two boojums exhibit the stick-slip dynamics at the six vertices of a hexagonal platelet (Fig. 2d,e), explaining the additional fine features in the temporal variations of angular speed. Moreover, the tendency of boojums to stick at vertices, substantially different from the random stick-slip behavior of spherical nematic colloids[19], shows how geometry of particles can be used to pre-define details of the rotational motion, potentially allowing for the control of the temporal evolution of instantaneous angular velocity at will.

Both clockwise and counterclockwise rotations of particles can be induced and robustly pre-selected by controlling the excitation light polarization **P**$_e$. For example, different linear polarizations of excitation light at angles $\theta$ measured with respect to **n**$_0$ prompt rotations of direction defined by the sign of $\theta$ (Fig. 5a). Polarizing optical micrographs reveal that details of **n(r)**-distortions around platelets depend on $\theta$ and slowly evolve with time (e.g. due to coarsening of small domains with initially different orientations of the azobenzene molecules and the LC in the vicinity of the platelets) (Fig. 6), also defining periodic features in the variations of angular speed of the rotating particle (Figs. 7). Once the rotational motion develops, angular speed can be further varied by changing $\theta$ (Fig. 5a), though the particle slows down for large deviations from the initial angle $\theta_i$ at which these dynamics are first prompted. This observation indicates the importance of the pre-history of the out-of-equilibrium director structure evolution in defining the exact details of rotational motion, as well as its complexity and sensitivity to numerous experimental details. Such out-of-equilibrium and history-dependent nature of this behavior poses a challenge in its quantitative modeling, though we shall see below that it can be understood invoking a simple model of interaction of light with a twisted director structure above and below the rotating platelet (Fig. 5c-e). Interestingly, despite of the relatively high particle rotation rates (~1 Hz), the relaxation of particle orientation upon turning off the blue-violet light is slow, >1000 s (Fig. 5b). This is consistent with previous observations[20] that azobenzene moieties within the monolayer assemblies in contact with the LC exhibit slow dynamics in dark and fast rotations in presence of blue-light activation, as well as with the fact that motion of these particles is highly overdamped, with negligible inertia effects.

**Diversity of shapes and pre-designed dynamics.** Periodic rotation and other types of dynamics can be achieved not only for the hexagonal platelets, but also for cogwheel-like (Fig. 8a-c and Supplementary Movie 6) and many other shapes, as well as various self-assemblies of particles (Fig. 8d,e and Supplementary Movie 7). Angular and temporal dependencies of the angular speed correlate with the geometric features and stick-slip motions of boojums on the surface of colloidal particles and their assemblies (Fig. 8), showing that shape of particles can be used to pre-engineer fine features of this out-of-equilibrium behavior and potentially the work done by ensuing colloidal motors. We envisage that the rotational dynamics of nematic colloids can be used to even further enrich their self-assembly and collective behavior, potentially giving origins to new forms of active matter and self-assembled micromachines.

**Modeling of the opto-mechanical feedback.** Our experiments (Figs. 1-8) reveal the emergent, history-dependent nature of colloidal spinning, which features stick-slip dynamics of singular boojums and discontinuous temporal evolution of director patterns, as well as complex behavior of polarization states of traversing light that interacts with the dynamically changing optical axis structures and slowly evolving monolayers of azobenzene molecules coupled to the rotating director. Precise modeling of such out-of-equilibrium effects in LCs is challenging because of the need of invoking description of the free energy costs due to gradients in the tensorial order parameter, anisotropic viscous and elastic properties, and also complex interaction of polarized light with the dynamic complex patterns of optical axis, which is beyond the scope of our present work and intended for future studies. However, our findings can be understood by modeling propagation of the excitation light through the central part of the platelet, with the assumption that the director field is uniformly twisted above and below it (Fig. 1). With these assumptions, Jones matrix modeling (see Methods and Fig. 5c-e) provides insights into how the polarization of excitation light evolves from linear to elliptical as it traverses through the twisted nematic slab (Fig. 1g), with the orientation of the long axis of the polarization ellipse also rotating, though slightly lagging the **n(r)** twist (Fig. 5c). This enables the well-defined coupling between orientation of the long axis of the polarization ellipse at the particle plane, **P**$_p$, and the orientation of azobenzene moieties within monolayers, which define the orientation of **n(r)** at platelet-LC interfaces. As the net twist of **n(r)** from the confining substrate to platelet across the cell increases, the ellipticity of light's polarization increases (Fig. 5c), reducing the strength of coupling between **P**$_p$ and the easy axis of molecular orientations and resulting in slippage of **n(r)** at the LC-platelet interfaces. This dynamics of **n(r)** establishes the feedback mechanism responsible for the light-driven colloidal motor. Fine details of this dynamics are revealed by monitoring temporal evolution of patterns of polarization state and intensity of light after passing the LC with oppositely twisted regions above and below the platelet (Fig. 2e,f) and are consistent with the results of our modeling (Fig. 5d,e). For example, the range of variation of experimentally characterized azimuthal angle and ellipticity (Fig. 2e,f) of elliptically polarized light traversing the cell through the middle of a platelet is consistent with the results of our modeling obtained for experimental material and sample parameters (Fig. 5c-e). Importantly, this modeling reveals how the opto-mechanical feedback mechanism of the motor results from the interaction of the dynamic twisted director structure with light and polarization-dependent interaction of light with the azobenzene monolayers and rotating platelets.

**Achieving translational motion.** In addition to rotation, low-intensity, unstructured light can also prompt translational displacements of colloids (Fig. 9). This combined translational and rotational dynamics is common for platelets with thickness of 1μm and larger (Fig. 9), whereas thin platelets with hexagonal shapes tend to exhibit rotational dynamics and Brownian-type translational diffusion (Figs. 1-8 and 10). The role of azobenzene molecules self-assembled at the edge faces of thin silica platelets characterized in the previous sections is negligible because the platelet thickness (≈0.5μm) is smaller than the surface anchoring extrapolation length $\xi=K/W$~1μm, where $K$ is the average Frank elastic constant and $W$ is the surface anchoring coefficient, so that the boundary conditions induced by azobenzene molecules are violated. For the thick ≥1μm platelets, the optical response of azobenzene at the edge faces further breaks symmetry of director distortions and causes conversion of rotational motion into translation. The translational motion can take different forms and be accompanied by various types of rotational dynamics, such as spinning or periodic angular vibrations of the platelet orientation (Fig. 9a,c and Supplementary Movies 8, 9). For example, for the particle with oscillating orientations, these observations are consistent with the time-averaged orientation of the platelet, which is tilted with respect to $\mathbf{n}_0$ and its equilibrium orientation without light exposure. This tilted oscillating orientation of a translating platelet can be understood as a dynamic equilibrium, for which edge faces tend to align so that the long axes of polarization ellipses of traversing light are roughly orthogonal to azobenzene molecules at the edge faces. The nonreciprocity of low-symmetry director deformations in the vicinity of edge faces of thick platelets during rotations and orientational oscillations causes the conversion of rotational motion into translational motion, in that resembling such processes in bacterial flagellum[14].

The speed of the observed translational motion typically ranges from several to tenths of μm/s, and the velocity is roughly along $\mathbf{n}_0$ when accompanied by angular vibrations of orientation (Fig. 9c,d) and at large angles to $\mathbf{n}_0$ when accompanied by spinning (Fig. 9a,b). This well-defined orientation of velocity vectors relative to $\mathbf{n}_0$ at the level of individual particles, caused by their alignment with respect to $\mathbf{n}_0$, is different from random directionality of individual particle motions observed in rotating colloidal systems activated by magnetic field[21], but resembles how individual bacteria[22,23] and topological solitons[24] move in well-defined directions in the background of a uniformly aligned LC. The linear translational velocity and linear velocity of the vertices of the hexagonal platelet exhibit similar periodic variations (Fig. 9e), indicating that rotations and translations of particles are indeed inter-related. Our findings demonstrate that the rotational dynamics we study, under suitable conditions, could be used to prompt translational motions driven by low-intensity light, effectively converting optical energy into motion on the individual colloidal particle basis. Although this is beyond the scope of our present study, we envisage that such optical energy-to-motion conversion could potentially enable novel light-powered active matter systems, with flocking[21], active turbulence and various other types of emergent collective out-of-equilibrium particle behavior[25, 26].

**Discussion.**

Light-driven rotation of our platelets is resisted by a viscous torque associated with an effective flow viscosity $\gamma_{\text{eff}}$ of the LC, which is analogous to rotations of colloidal particles in isotropic fluids[11], though also more complex because nematic colloids are accompanied by topological defects and elastic distortions that tend to increase the associated viscous drag[27, 28] and because full description of LC dynamics would generally require dealing with five independent flow viscosity coefficients[17] instead of $\gamma_{\text{eff}}$. Moreover, our particle spinning within the LC medium is additionally accompanied by large 0-180° twisting of **n(r)** above and below the platelet, which is very specific to LCs[17] and in fact similar to that found in LC displays[13], where electric fields drive rotation of **n(r)** and where these rotations are resisted by LC-specific torques due to orientational elasticity (related to the fact that full rotation of our light driven motor requires intensity above a certain threshold) and director's rotational viscosity $\gamma_{\mathbf{n}}$. When the studied LC is heated to isotropic phase, only the viscous torque associated with $\gamma_{\text{eff}}$ survives and resists particle rotations, though it also changes as compared to that in the nematic phase. Without particle present, dynamic rotations of **n(r)** are resisted by a viscous torque associated with $\gamma_{\mathbf{n}}$. In our case of spinning particle accompanied by twisting of **n(r)**, however, both these torques are present and comparable because $\gamma_{\text{eff}}$ and $\gamma_{\mathbf{n}}$ are of the same order of magnitude[17]. The torque associated with $\gamma_{\text{eff}}$ can be characterized by probing rotational diffusion of the platelets in response to thermal fluctuations[29] when such platelets are immersed in the LC at no optical torques and for uniform director across the LC cell. On the other hand, the torque associated with the dynamics of $\mathbf{n}(\mathbf{r})$ can be estimated using known value $\gamma_\mathbf{n}$=0.08 Pa s of the rotational viscosity of the used LC[30]. The power of mechanical work associated with light-driven rotational dynamics of the platelet and $\mathbf{n}(\mathbf{r})$ around it is then estimated as $W_m \approx \omega^2(k_B T/D_\varphi + \gamma_\mathbf{n} d A_p)$, where $\omega$ is the angular frequency of particle rotation, $d$ is the sample thickness, $A_p$ is the area of the platelet's hexagonal face, $k_B$=1.38×10$^{-23}$ J K$^{-1}$ is the Boltzmann constant, $T$=298 K is temperature and $D_\varphi$=3.8×10$^{-4}$ rad$^2$ s$^{-1}$ is the rotational diffusion coefficient of particles obtained from probing their Brownian motion (see Methods). About 0.5% of the excitation power is absorbed by azobenzene molecular monolayers, which yields an estimate of the optical-to-mechanical power conversion efficiency of $\eta \approx 10^{-3}$% at 200 nW of excitation light and the corresponding rotation frequency of $\approx$ 1 Hz (Fig. 2b). The scaling of rotation frequency with optical power $W_o$ above its threshold value $W_{oth}$, $\omega \propto (W_o - W_{oth})^{1/2}$, is qualitatively consistent with experimental measurements (Fig. 2b). The threshold optical power $W_{oth}$ needed to prompt continuous rotation is associated with overcoming the energetic costs of the twist elastic distortions in $\mathbf{n}(\mathbf{r})$ and prompting the stick-slip dynamics of boojums. Ambient sunlight intensity ~50 mW cm$^{-2}$ on a sunny day is significantly higher than the intensity of 8 mW cm$^{-2}$ corresponding to $W_{oth}$ measured in our experiments, indicating that such colloidal motors can be driven simply by harnessing solar energy and ambient white light (Figs. 3). Although the efficiency of our rotary motor is so far nowhere close to that of some biological motors, like the near-100%-efficient flagellum motor with ~1kHz rotations and other biological motor marvels[14], it is superior to that of other man-made light-driven colloidal motors (to the best of our knowledge, the highest efficiency reported for light-driven man-made micromotors[31, 32] is ~10$^{-8}$ %, 5 orders of magnitude lower than ~10$^{-3}$ % achieved here).

To conclude, self-assembly on molecular and colloidal scales yields a light-driven rotary motor within which cooperative responses of hundreds of millions of dye molecules and trillions of LC molecules interact with a colloidal platelet to convert optical power into mechanical work. We have uncovered the LC-based feedback mechanism enabling realization of different forms of colloidal dynamics, ranging from periodic oscillations to continuous unidirectional rotation, as well as demonstrated that light intensity and geometric shape of particles can be used in controlling angular speed of particles. We envisage that collective behavior of our light-driven motors can yield new forms of active matter[25, 26] and that they can be self-assembled into colloidal micro-machines, of interest for many technological applications[3-7].

**Methods.**

**Sample preparation.** The details of our experimental system are presented in Fig. 10. The used photo-responsive azobenzene-containing molecules (Fig. 10f) were synthesized by following well-established protocols[33]. In 20 mL of dichloromethane, 1.35 g of methyl red and 1.11 g of 1,3-dicyclohexylcarbodiimide were dissolved and then combined with 1.19 mL of (3-aminopropyl) triethoxysilane under nitrogen. After reacting overnight under stirring, the solution was purified by column chromatography (50% ethyl acetate, 50% hexane) and dried to yield dark red crystals of the desired azobenzene-containing dye with chemical structure shown in Fig. 10f and the corresponding absorbance spectra given in Fig. 10h. Silica micro-platelets (Fig. 10b-e) were fabricated via direct laser writing photolithography[34]. First, plasma-enhanced chemical vapor deposition was used to deposit a layer of silica of 0.5 μm or more in thickness on a silicon wafer followed by spin-coating an additional layer of photoresist AZ5214 (from Clariant AG). Geometric shapes of particles were defined in the photoresist layer with a direct laser-writing system DWL 66FS (Heidelberg Instruments) operating at 405 nm and subsequently in the silica layer by inductively coupled plasma etching. The photoresist mask was then removed with acetone and the silicon substrate was further etched using selective inductively coupled plasma to detach micro-platelets from the substrate. To facilitate surface binding of azobenzene-containing dye molecules, the platelets were immersed in a piranha solution (50% of 30 wt% hydrogen peroxide, 50% of 98 wt% sulfuric acid by volume) for 1 h to produce hydroxylated surfaces. They were then submerged in 20 mL toluene solution of 5 mg of the azobenzene-containing dye and 20 μL n-butylamine at elevated temperature of 45 °C overnight, followed by rinsing with toluene and curing at 120 °C for 2 hrs. Ultrasonic bath of the resulting substrate yielded a suspension of micro-platelets in isopropanol with self-assembled monolayers of azobenzene-containing dye atop of particles, which was then mixed with a nematic LC pentylcyanobiphenyl (5CB, obtained from Frinton Laboratories, Inc.). After isopropanol evaporated at room temperature, the LC colloidal dispersion was infiltrated between glass plates, with gap defined by glass spacers and boundary condition by unidirectional rubbing of polyimide PI2555 (HD Microsystem) spin-coated on the glass plates. Unless noted otherwise, all chemicals were purchased through Sigma-Aldrich and used as received.

**Experimental techniques.** Rotational motions of particles and **n(r)** were imaged and characterized with polarizing optical microscopy based on an upright microscope (BX51, Olympus) (Fig. 10a). To avoid undesired excitation, experiments were done in a dark room and a red filter (FF01-640/14-25, Semrock) was placed before the microscope's condenser to block violet-blue component of imaging light during polarizing microscopy observations of particle rotation in the LC (Fig. 10a). Optical images and videos were recorded by a charge-coupled device camera (Grasshopper3, PointGrey) and analyzed with ImageJ and its plugins (freeware from the National Institute of Health) to extract details of the rotational motion of colloidal particles. The normalized spectra of the used microscope's white light, blue excitation light and red imaging light under above described conditions of experiments are shown in Fig. 10g, which are designed to optimize or avoid response of the azobenzene molecules (see the absorption spectrum in Fig. 10h) to excitation or imaging light, respectively. The red-light videomicroscopy also allowed for the characterization of rotational (Fig. 10i) and anisotropic translational diffusion of colloidal platelets (Fig. 10j) under conditions of avoiding excitation of azobenzene molecules and no spinning.

Characterization of patterns of director field and optical axis of birefringent materials often benefits from probing polarization states of light passing through these samples[35]. To quantitatively probe spatial patterns of polarization ellipse for initially linearly polarized light after it passes through the LC with a rotating platelet, an additional broadband quarter-wave plate (AQWP05M-600, Thorlabs) was inserted before the analyzer and rotated to obtain images at different angles. Polarizing optical micrographs were obtained for different waveplate orientations every 22.5°, starting with the orientation of the waveplate's fast axis parallel to $n_0$ and the analyzer. These images were then numerically processed to produce visualizations of polarization ellipses[36] on the pixel-by-pixel basis that characterize light's polarization state after it passes through the sample (Fig. 2f). Three-photon excitation fluorescence microscopy[37] was used to characterize three-dimensional director structures and the location of colloidal particles within the LC samples (Fig. 10d,e).

A mercury lamp (U-LH100HG, Olympus) in the reflection channel of the BX51 Olympus microscope (Fig. 10a) was used as a source of the blue excitation light, with the excitation wavelength range defined by a band pass filter HQ480/40X (Chroma) (Fig. 10g). Neutral density filters and a rotatable linear polarizer were also inserted into the excitation optical path to control light intensity and polarization (Fig. 10a). Objectives of different magnification (20×-60×, from Olympus) were used to project blue light onto and collect transmitted light from the sample. Optical power of the excitation light delivered per particle was characterized with a photodiode (PDA100A, from Thorlabs) while accounting for the relative areas of a platelet and the microscope's field of view.

**Opto-mechanical feedback and polarimetric imaging.** Key to the opto-mechanical feedback mechanism described in this work is the propagation of light through a twisted structure of the director field, as depicted in Fig. 1g. In order to get insights into the relations between polarization of traversing light and director structure, we model propagation of light through the middle of the colloidal platelets and twisted LC above and below it. For this, we write Maxwell's equations in a matrix form within an approach known as the Jones matrix method[13]. The polarization state of light traversing through the twisted LC slab can be then expressed as:

$$\begin{pmatrix} V'_e \\ V'_o \end{pmatrix} = \begin{pmatrix} \cos X - i\frac{\Gamma \sin X}{2X} & \phi\frac{\sin X}{X} \\ -\phi\frac{\sin X}{X} & \cos X + i\frac{\Gamma \sin X}{2X} \end{pmatrix} \begin{pmatrix} V_e \\ V_o \end{pmatrix} \quad (1)$$

where $\begin{pmatrix} V'_e \\ V'_o \end{pmatrix}$ and $\begin{pmatrix} V_e \\ V_o \end{pmatrix}$ are Jones vectors of the outgoing and incoming light in the local principal coordinate system defined along and perpendicular to the optical axis of LC, which is along **n(r)**. Taking into account that the blue excitation light traversing the platelet passes through a slab of LC twist, which is approximately half of the cell thickness, we define the following variables describing the light-LC interaction at the moment of reaching the colloidal platelet in the cell midplane:

$$\Gamma = \frac{\pi(n_e - n_o)d}{\lambda} \tag{2}$$

$$X = \sqrt{\phi^2 + \left(\frac{\Gamma}{2}\right)^2} \tag{3}$$

where $\phi$ is the director twist angle, $d$ is the thickness of the LC cell (defining the thickness of the twisted LC slab through which light propagates before reaching the platelet as $\approx d/2$), $\lambda$ is the wavelength of the light and $n_e$-$n_o$ is the difference between extraordinary and ordinary indices (birefringence) of the LC. When the polarization of the incident light is parallel to the far-field director $\mathbf{n}_0$, $\theta = 0°$, its Jones vector is

$$\begin{pmatrix} V_e \\ V_o \end{pmatrix} = \begin{pmatrix} 1 \\ 0 \end{pmatrix} \tag{4}$$

and a simple matrix calculation yields

$$\begin{pmatrix} V'_e \\ V'_o \end{pmatrix} = \begin{pmatrix} \cos X - i\frac{\Gamma \sin X}{2X} \\ \phi \frac{\sin X}{X} \end{pmatrix} \tag{5}$$

, which corresponds to a polarization state with ellipticity and long-axis azimuthal orientation given by

$$e = \tan\left(\frac{1}{2}\sin^{-1}\left(\frac{\Gamma \phi}{X^2}\sin^2 X\right)\right) \tag{6}$$

$$\tan 2\psi = \frac{2\phi X \tan X}{(\phi^2 - \Gamma^2/4)\tan^2 X - X^2} \tag{7}$$

Using experimental parameters $\lambda$=490 nm, $n_e$-$n_o$=0.2, and $d/2$=2.4 μm, we characterize ellipticity $e$ and azimuthal orientation $\theta = \psi + \phi$ versus the director twist angle (Fig. 5c), revealing details of how the twist structure changes the polarization state of the excitation light.

Polarizing optical micrographs and polarimetric textures of a sample with a platelet in the LC (Fig. 2) obtained while it rotates can be explained by exploring the polarization state of the imaging light at the exit of the cell. The Jones matrix of light propagation through LC twist as a function of director twist angle can be expressed as

$$M(\phi) = \begin{pmatrix} \cos X - i\dfrac{\Gamma \sin X}{2X} & \phi \dfrac{\sin X}{X} \\ -\phi \dfrac{\sin X}{X} & \cos X + i\dfrac{\Gamma \sin X}{2X} \end{pmatrix} \quad (8)$$

In our polarized imaging experiments (Fig. 2f), the polarization state of imaging light is probed after it passes through two $d/2$–thick slabs of LC twist, one above and one below the platelet (Fig. 1g). We assume that the director twist angles in the two slabs are the same by magnitude but have opposite signs. The combined Jones light propagation matrix is simply $M(-\phi) \cdot M(\phi)$, so that the polarization state of outgoing light can be found as

$$\begin{pmatrix} V''_e \\ V''_o \end{pmatrix} = M(-\phi) \cdot M(\phi) \cdot \begin{pmatrix} V_e \\ V_o \end{pmatrix} \quad (9)$$

Given the same linear polarization of incident light as above, the Jones vector of the outgoing imaging light is

$$\begin{pmatrix} V''_e \\ V''_o \end{pmatrix} = \begin{pmatrix} \dfrac{\phi^2 \sin^2 X}{X^2} + \left(\cos X - \dfrac{i\Gamma \sin X}{2X}\right)^2 \\ -\dfrac{i\phi\Gamma \sin^2 X}{X^2} \end{pmatrix} \quad (10)$$

The corresponding ellipticity and azimuthal orientation angle can be expressed as

$$e = \tan\left(\dfrac{1}{2}\sin^{-1}\left(2\dfrac{\Gamma\phi}{X^2}\sin^2 X \left(1 - 2\sin^2 X + 2\dfrac{\phi^2 \sin^2 X}{X^2}\right)\right)\right) \quad (11)$$

$$\tan 2\theta = \dfrac{2\Gamma^2 \phi X \sin^3 X \cos X}{X^4 - 2\phi^2 \Gamma^2 \sin^4 X} \quad (12)$$

Additionally, the light transmission coefficient for each pixel of the polarizing optical micrograph obtained under crossed polarizers reads

$$T = |V''_o|^2 = \dfrac{\phi^2 \Gamma^2 \sin^4 X}{X^4}, \quad (13)$$

and is found as $1 - T$ for parallel polarizers. Using experimental parameters and $\lambda$=640 nm, we characterize the polarization states of light passing through the LC cell with a rotating platelet (Fig. 5d,e), finding these results consistent with the experimental characterization (Fig. 2).

**Light-controlled boundary conditions and elastic distortions.** To model interaction of our LC colloidal motors with light, we assume infinitely strong (fixed) surface boundary conditions at the confining substrates of the LC cell. The free energy potential describing the generally soft surface boundary conditions at the LC-platelet interfaces is dependent on light's intensity $I$ and its polarization state and can be expressed as

$$F_{platelet} = [\sigma(I,e)/2] \int (\mathbf{P}_p \cdot \mathbf{n}_p)^2 dS \tag{14}$$

,where $\sigma(I,e)$ is the surface anchoring coefficient (an anisotropic part of surface tension) and $\mathbf{n}_p$ defines the orientation of $\mathbf{n(r)}$ at the platelet's surface. This free energy term is always minimized when $\mathbf{P}_p \perp \mathbf{n}_p$, as defined by the response of azobenzene-containing monolayers, but the strength of such coupling through light-directed boundary conditions depends on both intensity and polarization state of light traversing the platelet, as determined by $\sigma(I,e)$. When polarization state changes so that the long axis of ellipse $\mathbf{P}_r$ rotates, this prompts rotation of $\mathbf{n}_p$ and rotation of both the platelet and $\mathbf{n(r)}$ in the LC bulk, though slipping occurs at large $\epsilon$ and low $I$ due to weak coupling defined by the ensuing small $\sigma(I,e)$. The rotation-driven $\mathbf{n(r)}$ distortions in the LC bulk are resisted by the elastic free energy describing the energetic cost of these distortions at fixed boundary conditions at the cell's confining substrates

$$F_{elastic} = \int \left\{ \begin{array}{c} \frac{K_{11}}{2}(\nabla \cdot \mathbf{n})^2 + \frac{K_{22}}{2}[\mathbf{n} \cdot (\nabla \times \mathbf{n})]^2 + \frac{K_{33}}{2}[\mathbf{n} \times (\nabla \times \mathbf{n})]^2 \\ -K_{24}[\nabla \cdot [\mathbf{n}(\nabla \cdot \mathbf{n}) + \mathbf{n} \times (\nabla \times \mathbf{n})]] \end{array} \right\} dV \tag{15}$$

where $K_{11}$, $K_{22}$, $K_{33}$, and $K_{24}$ are Frank elastic constants for splay, twist, bend and saddle splay director deformations, respectively. The competition of corresponding light-dependent surface anchoring free energy term and bulk elastic free energy defines conditions for slipping of **n(r)** at the platelet-LC interfaces, which, in turn, defines the dynamics of our system.

**Data availability** All datasets generated and analyzed during the current study are available from the corresponding author on reasonable request.

**Supplementary Information** is available in the online version of the paper.

**Acknowledgements.** This research was supported by the National Science Foundation Grant DMR-1810513. We thank Bohdan Senyuk and Slobodan Zumer for discussions, Sungoh Park for assistance with scanning electron microscopy imaging, Bohdan Senyuk for assistance with polarimetric imaging and Dawei Zhang for assistance with material purification.

**Author contributions.** Y.Y. and Q.L. synthesized azobenzene molecules. Q.L. fabricated microparticles. Y.Y. and G.N.A. performed experiments and analyzed data. I.I.S. conceived and designed the project, as well as wrote the manuscript with feedback from all authors.

**Additional information**. Reprints and permissions information is available at www.nature.com/reprints. The authors declare no competing financial interests. Readers are welcome to comment on the online version of this article at www.nature.com/nature. Correspondence and requests for materials should be addressed to I.I.S. (ivan.smalyukh@colorado.edu).

**Competing interests.** The authors declare no competing interests.

# Figures:

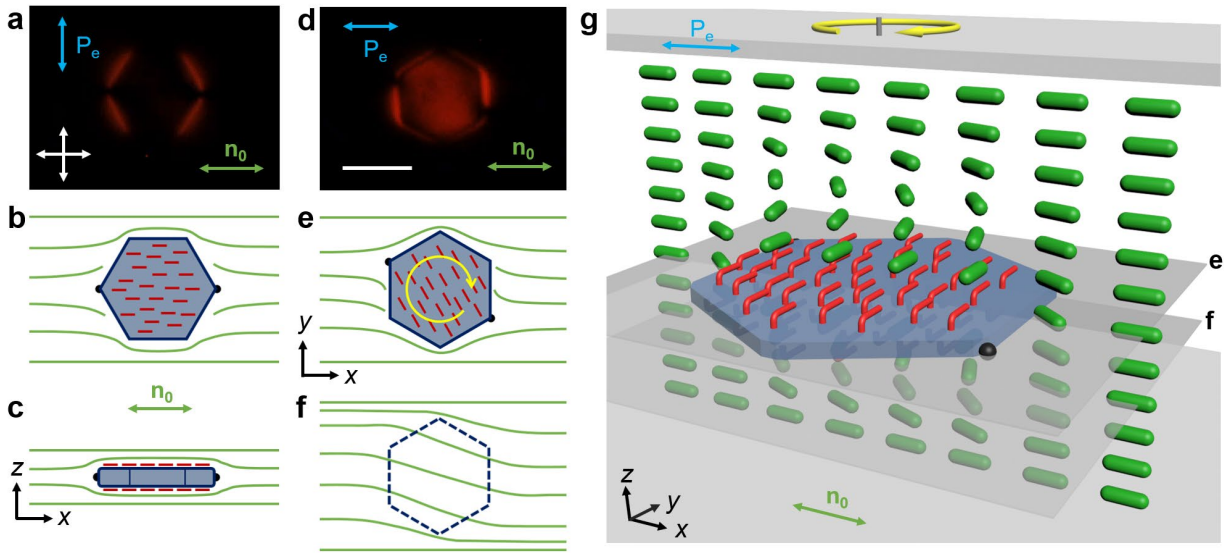

**Figure 1 | Colloidal platelet with self-assembled azobenzene monolayers in a nematic liquid crystal. a,** polarizing optical micrograph obtained under red imaging light when the platelet is illuminated by linearly polarized blue light with $\mathbf{P}_e \perp \mathbf{n}_0$ (green double arrows). $\mathbf{P}_e$ is indicated by a blue double arrow and crossed polarizers of the microscope are shown using white double arrows. **b,c,** corresponding schematics of $\mathbf{n}(\mathbf{r})$ (green lines) and orientations of trans-state azobenzene moieties (red rods) within the self-assembled monolayers **b**, in the plane parallel to cell substrates (*xy* plane) and **c**, in the cell's vertical cross-section (*xz* plane). Black filled hemi-circles at the vertices of platelet along $\mathbf{n}_0$ depict the boojum defects. **d,** polarizing micrograph of the same particle as in **a**, but when it rotates under blue-light illumination with $\mathbf{P}_e \parallel \mathbf{n}_0$. Scale bar is 5 μm. **e,f,** Schematics of $\mathbf{n}(\mathbf{r})$ and azobenzene orientations in the plane of platelets (**e**), and in the plane beneath it (**f**), both marked in **g**; the circular yellow arrow shows handedness of rotation. **g,** Three-dimensional schematic of the self-assembled molecular-colloidal light-driven motor between two confining glass plates, with the green rods represent LC molecule and the azobenzene molecules of self-assembled monolayers shown in red.

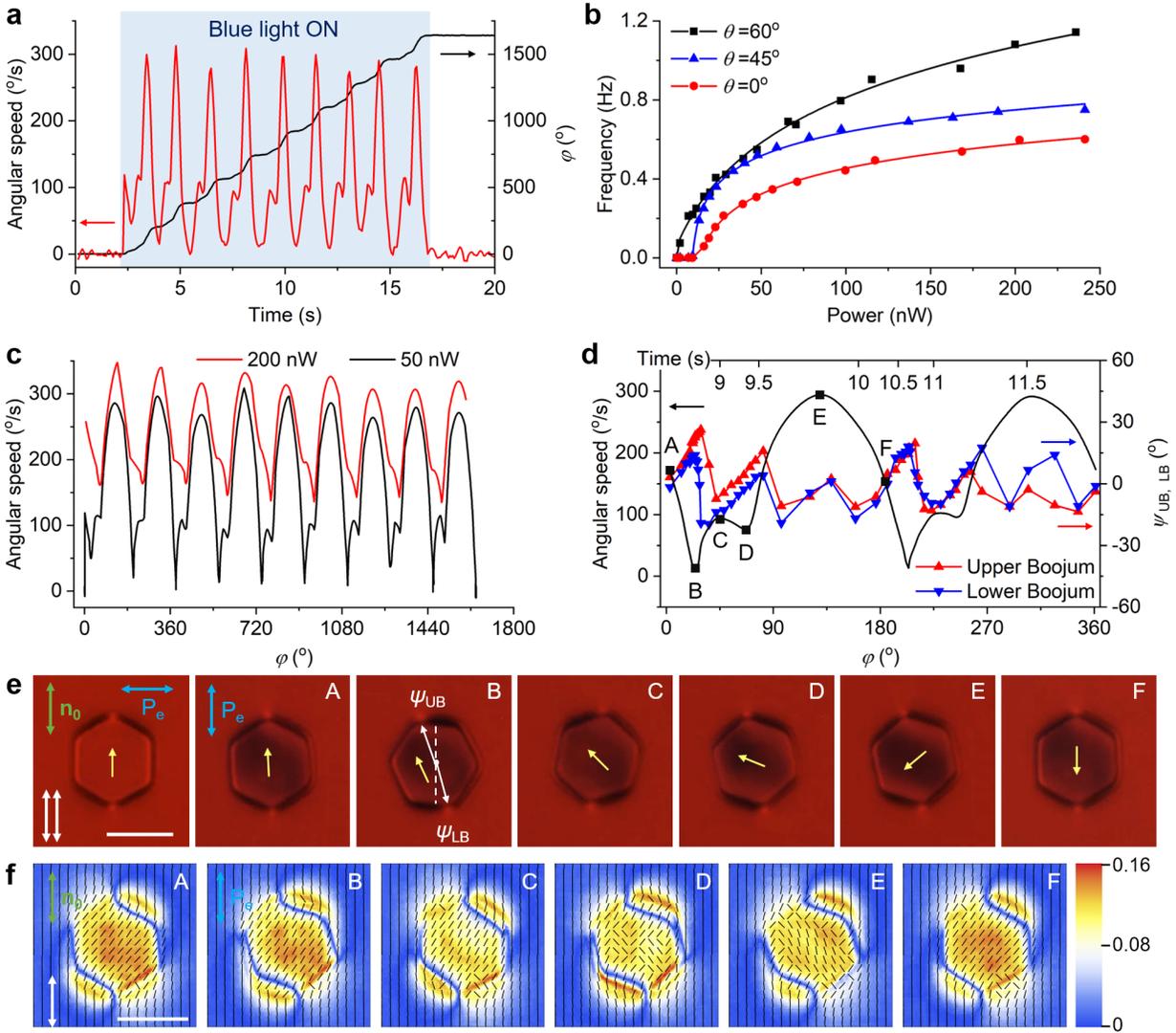

**Figure 2 | Characterization of light-driven colloidal spinning motion. a**, Angular speed and the net azimuthal rotation angle $\varphi$ versus time probed upon turning the blue excitation light on and then off. **b**, Frequency of rotation versus optical power of light incident on the colloidal platelet at different angles $\theta$ between $\mathbf{P}_e$ and $\mathbf{n}_0$. Symbols represent experimental measurements and solid lines are for eye guiding. **c**, Angular speed versus $\varphi$ at different powers of incident light. **d**, Angular speed versus $\varphi$ within a single full rotation compared to respective angular variations of angles $\psi_{UB}$ and $\psi_{LB}$ characterizing orientations of vectors that connect particle's center of mass with upper and lower boojums, respectively, as marked in (**e**). **e**, Polarizing optical micrographs, obtained for samples placed in-between parallel linear polarizers along $\mathbf{n}_0$, with the angular positions (shown by yellow arrows) correlated with different angular positions in the dependencies shown in **d**. **f**, Spatial patterns of polarization state of the red imaging light after passing through the sample. Thin black rods indicate orientation of the long axes of the polarization ellipses and background color ranging from blue to red represents distribution of the absolute value of ellipticity of the polarization ellipse. Labels A-F correlate these polarization ellipse data to the polarizing micrographs at the same angular positions shown in **e**. Incident imaging light is polarized along $\mathbf{n}_0$ as shown by the white double arrow. Scale bars are 5 μm.

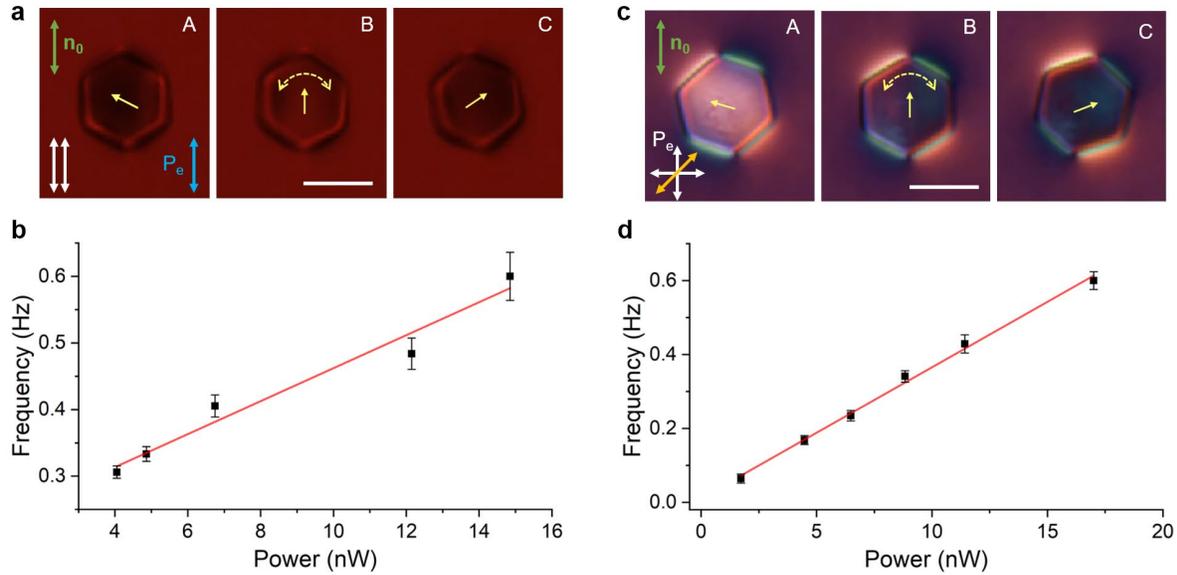

**Figure 3 | Light-driven periodic angular oscillations of colloidal particles. a**, frames from a polarizing optical microscopy video (Supplementary Movie 4) showing oscillation of a micro-platelet within an angular range < 180° in a 7 μm cell at low intensity of blue excitation light. Frames A and C show the micro-platelet at the two extreme angular displacements. Images were taken under parallel polarizers (white double arrows). Polarization direction of the blue excitation light (marked with a blue double arrow) is along the far-filed director $\mathbf{n}_0$. **b**, Oscillation frequency as a function of the optical excitation power of blue light delivered to such a particle. **c**, Frames from a polarizing microscopy video (Supplementary Movie 5) showing oscillation of a micro-platelet in a 4 μm cell under white light illumination. Frames A and C depict the micro-platelet at the extreme angular displacement. Images were taken using crossed polarizer and analyzer in the polarizing microscope, with an additional 530 nm retardation plate inserted between polarizers after the sample. Orientation of crossed polarizers and the quarter-wave plate's slow axis are shown with crossed white arrows and a yellow double arrow, respectively. Note that white light was used here as both the imaging and excitation light source and its polarization $\mathbf{P}_e$ was also set to be along $\mathbf{n}_0$. **d**, Oscillation frequency as a function of the optical power of white light. Scale bars are 5 μm. Data points in **b** and **d** are mean ± s.e.m; red lines are for eye guiding.

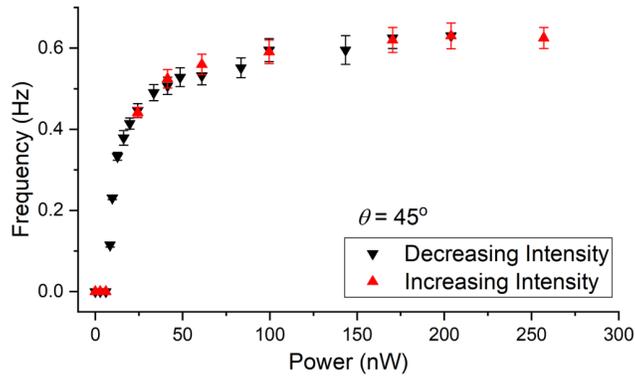

**Figure 4 | Comparison of rotation frequency dependences on optical power obtained while decreasing and increasing intensity.** Rotation frequency was measured first on a platelet spinning under blue excitation light of decreasing intensity and then, using the same particle, upon increasing intensity. These data were obtained for $\mathbf{P}_e$ at $\theta=45°$ with respect to the far-field director $\mathbf{n}_0$. Data points are mean ± s.e.m.

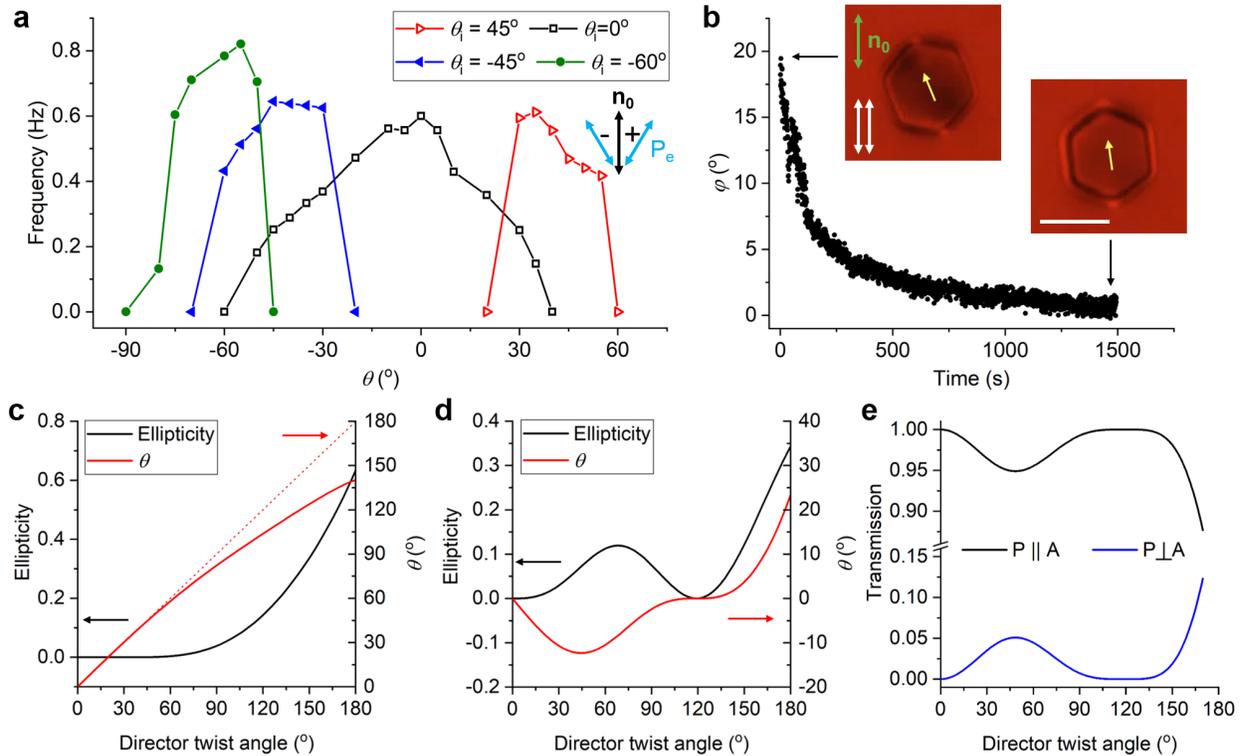

**Figure 5 | Control, relaxation and polarization dependence of colloidal spinning. a,** Demonstration of robust selection of handedness of colloidal spinning by illuminating the platelet with blue light linearly polarized at positive (prompts clockwise spinning of platelets) or negative (counterclockwise spinning) initial angles $\theta_i$ between $\mathbf{P}_e$ and $\mathbf{n}_0$, as well as further control of rotation frequency by then rotating $\mathbf{P}_e$. Definition of the sign of $\theta$ is shown in the inset. **b**, Slow relaxation of azimuthal orientation angle $\varphi$ of the platelet relative to its equilibrium position upon switching off the blue-light excitation. The insets show the representative polarizing optical micrographs. Scale bar is 5 μm. **c**, Theoretical prediction for the polarization state (ellipticity and azimuthal orientation of the long axis of the polarization ellipse $\theta$) of the blue excitation light when it reaches the platelet after passing the LC slab of $\approx d/2$ thickness with the amount of director twist quantified by the net twist angle. The rotation of the long axis of the polarization ellipse is lagging the rotation of director within the twisted structure (shown with the red dashed line for comparison). **d**, Calculated polarization state of the red imaging light when it exits the LC cell in the geometry shown in Fig. 1g. In the calculation, the red imaging light is assumed to pass through the LC slab above the platelet and through the one below the platelet, both with the same $d/2$ thickness and magnitude (though opposite signs) of director twist. **e**, Calculated transmission coefficient of the red imaging light under parallel (P ∥ A) and crossed (P ⊥ A) polarizers as a function of the director twist angle (by absolute value) within each of the LC slabs. The following experimental parameters were used in these calculations: wavelength of blue excitation light 490 nm, wavelength of red imaging light 640 nm, LC birefringence $\Delta n=0.2$, and thickness of the LC slab $d/2=2.4$ μm. The linear polarization of incident excitation or imaging light was set to be parallel to $\mathbf{n}_0$ ($\theta=0°$). Values in plots **c**, **d**, and **e** are calculated based on Jones matrix modeling of light propagation in a LC cell.

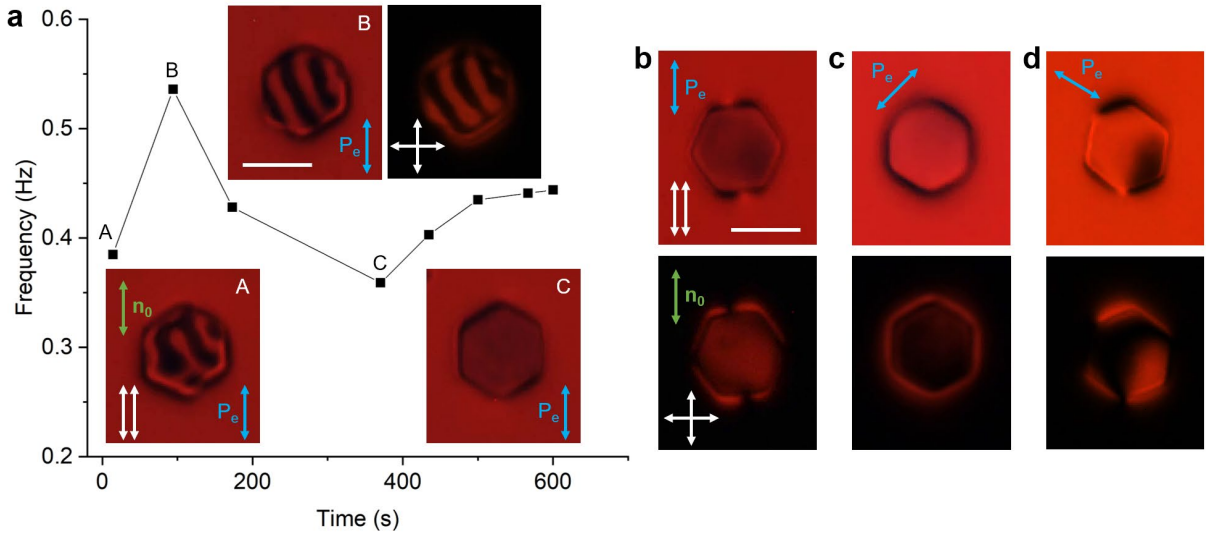

**Figure 6 | Long-term temporal evolution of particle dynamics and polarizing textures. a,** frequency of rotation versus time as a platelet spins upon shining blue light with polarization parallel to $n_0$. At the onset of rotation, micro-platelets typically have polarizing optical textures with alternating bright and dark patches/stripes (insets A and B), which is due to domains of varying azobenzene moiety orientation corresponding to different amounts of twist within the LC above and below the platelet. After some time, the photo-sensitive molecules synchronize their orientation within different domains while reducing elastic energy within the LC, producing more uniform patterns (see an example in the inset C). The linear polarization directions of the blue light are kept along $n_0$, as shown using a blue double arrow. The plot shows changes of rotation frequency over time. Data points labelled A-C corresponds to the inset images A-C. Due to changing azobenzene orientation domains and director twist, the rotation frequency fluctuates with the evolution of textures at the beginning, but is stabilized after the "synchronization" corresponding to the monodomain-like orientation of azobenzene molecules and uniform twist above and below the platelet. **b**, boojums moving along the edges of platelets, clearly visible under polarizing optical microscope when obtaining micrographs with (top) parallel and (bottom) orthogonal orientations of the polarizer and analyzer. **c, d**, frames from movies that illustrate textures of director field during the platelet spinning when induced with linear polarization of blue light at 45° (**c**) and -60° (**d**) with respect to $n_0$. Scale bars are 5 μm.

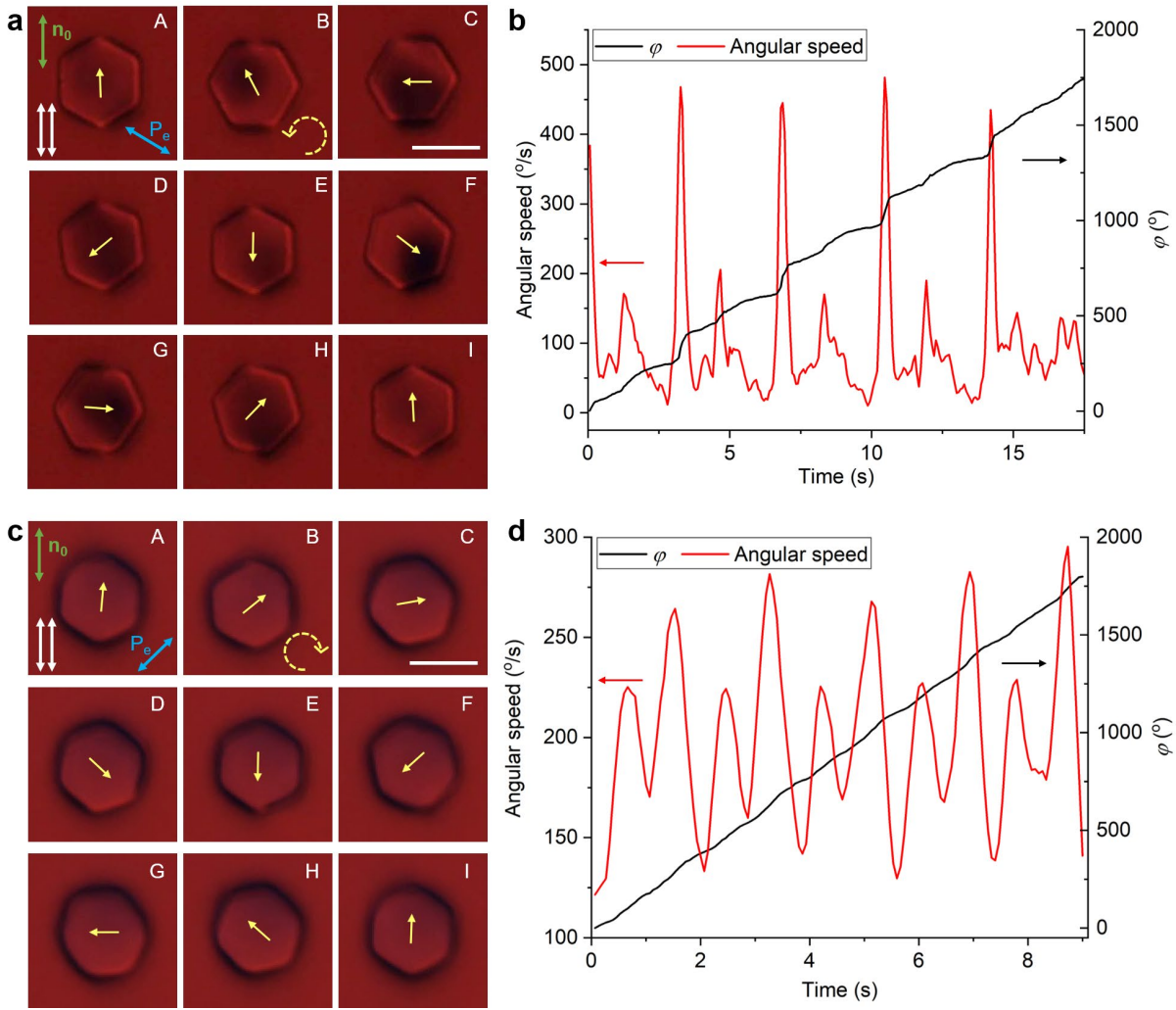

**Figure 7 | Particle spinning for polarizations of excitation light at $\theta \approx -60°$ and $\theta \approx 45°$. a, c,** Sequences of images showing a micro-platelet at different orientations during one rotation period under polarization of excitation light at $\theta \approx -60°$ (**a**) and $\theta \approx 45°$ (**c**). Polarizing optical images were taken under parallel polarizers shown with two adjacent white double arrows. Polarization direction of the blue excitation light is shown using a blue double arrow. The particle rotates counterclockwise (**a**) and clockwise (**c**), as indicated by the yellow dashed arrow. Its angular positions in each image are shown using yellow arrows. Scale bars are 5 µm. **b, d,** The net azimuthal rotation angle $\varphi$ (black curve) and angular speed (red curve) versus time during five periods of platelet rotation corresponding to platelet rotation shown in **a** and **c**, respectively.

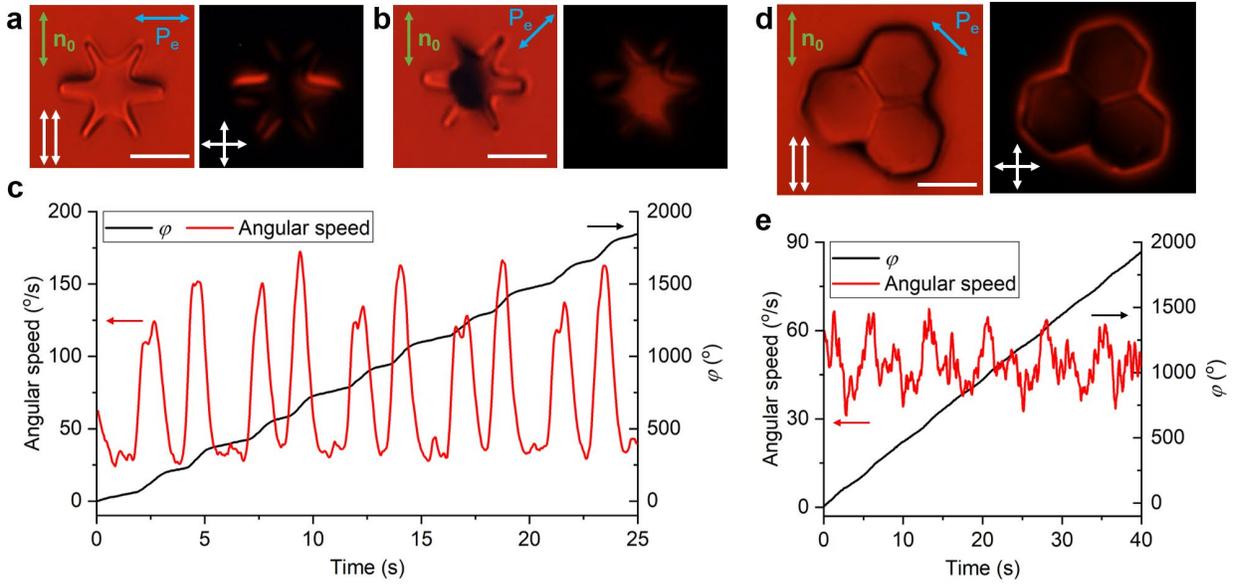

**Figure 8 | Spinning of a cogwheel-like particle and self-assembled hexagonal platelets. a, b,** Polarizing optical micrographs obtained with red imaging light and under linearly polarized blue light illumination $P_e$ perpendicular to $n_0$ (**a**), and at 45º to $n_0$ (**b**), when platelet rotates. **c,** Corresponding angular speed versus time for $P_e$ at 45º to $n_0$. **d,** Polarizing optical micrograph obtained with red imaging light and under linearly polarized blue light with $P_e$ at -45º to $n_0$. **e,** Corresponding angular speed and azimuthal orientation angle $\varphi$ versus time. Scale bars are 5 μm.

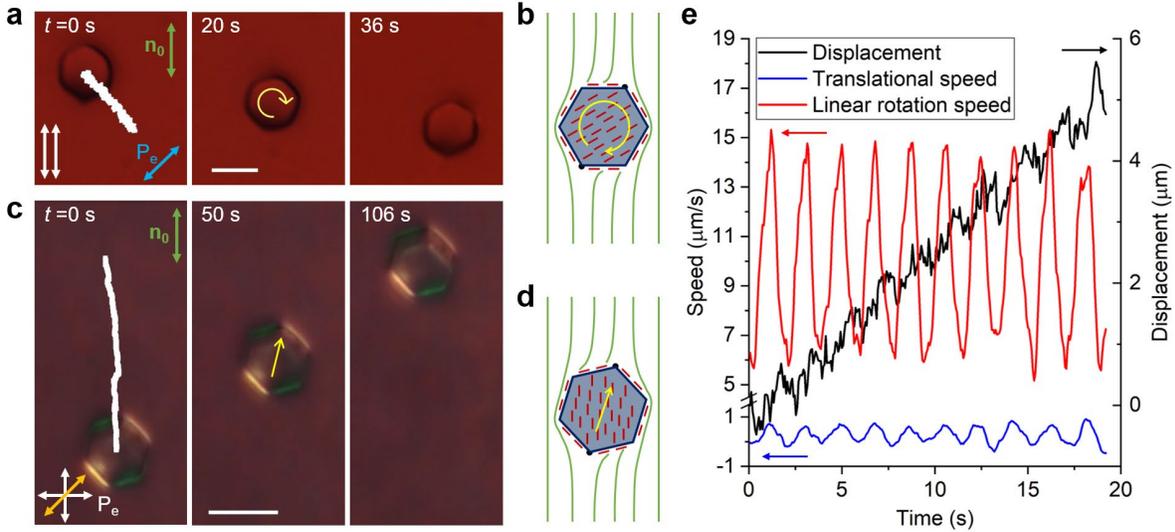

**Figure 9 | Translational motion of ≈1 μm thick colloidal platelets powered by light. a**, Snapshots of translational motion of a platelet accompanied by its spinning (Supplementary Movie 8) under blue excitation light with polarization $\theta \approx 45°$ (blue double arrow). Elapsed time is marked on images obtained under microscope's parallel polarizers shown with two adjacent white double arrows along $\mathbf{n}_0$. **b**, Schematic of director $\mathbf{n}(\mathbf{r})$ (green lines) around the platelet with azobenzene molecules (red rods) on its large-area top and bottom surfaces and along the edge faces. Yellow circular arrows in **a** and **b** show direction of rotation. **c**, Snapshots of translational motion of a platelet accompanied by periodic oscillations of orientation (Supplementary Movie 9) under polarized white light with polarization at $\theta \approx 90°$ with respect to $\mathbf{n}_0$. The symmetry axis (yellow arrow) of the platelet tilts away from the far-field director at about 13° during this motion. Images are shown for elapsed times marked on images obtained under crossed polarizers (white double arrows) with retardation plate (slow axis indicated by orange double arrow). The far-field director $\mathbf{n}_0$ is shown using green double arrows and scale bars are 5 μm. **d**, Schematic of $\mathbf{n}(\mathbf{r})$ and platelet with a self-assembled monolayer of azobenzene molecules corresponding to **c**. **e**, Time dependence of translational displacement (black line) and speed (blue line) of the platelet shown in **a**. The speed of particle translation exhibits the same periodicity as the linear speed (red line) of the hexagon's vertices with respect to its center, indication that the translational dynamics is directly coupled to angular rotation.

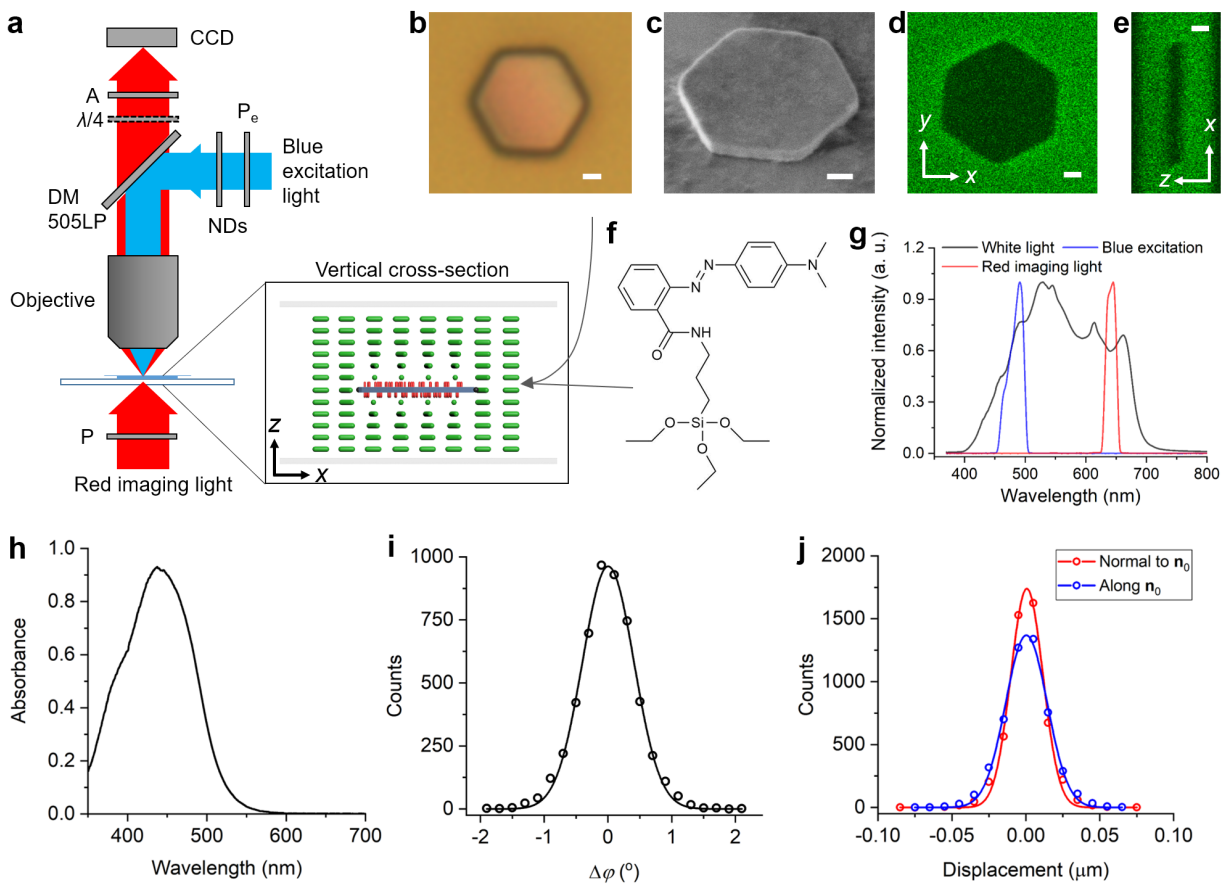

**Figure 10 | Experimental characterization of platelets with azobenzene monolayers. a**, An integrated imaging and excitation setup. White light from a mercury lamp is filtered to provide blue excitation light while red light is used for transmission-mode polarizing optical microscopy. $P_e$ is a rotatable polarizer used to define linear polarization direction of the blue excitation light; NDs are neutral density filters used to control excitation power. DM is a dichroic mirror reflecting the blue excitation light but transmitting the red imaging light. P and A are polarizer and analyzer of the polarizing microscope, respectively. $\lambda/4$ is a broadband quarter-wave plate used for polarimetric imaging. Inset shows a schematic of a cell with a particle. **b**, Optical micrograph of a platelet on a substrate obtained without polarizers. **c**, Scanning electron microscopy image showing that the silica platelets used for probing rotational dynamics are about 0.5 μm in thickness. **d,e** Three-photon excitation fluorescence microscopy[37] images of a cell with a platelet in its midplane obtained (**d**), for the cross-sectional plane passing through the platelet's middle and (**e**), for the vertical plane orthogonal to the plane of cell and platelet's large area faces, passing through the platelet's center of mass. **f**, Chemical structure of the used derivative methyl red molecule containing azobenzene moiety. **g**, Normalized spectra of the red and white imaging light and blue excitation incident light used in experiments. Scale bars are 1 μm. **h**, Absorbance spectrum of the used photosensitive molecules of derivative methyl red in toluene at concentration of $5 \times 10^{-5}$ M. The spectrum was obtained using a 1-cm-thick cuvette and a spectrometer Cary 500 (from Varian). **i**, A histogram quantifying the angular diffusion of a platelet in a plane containing $\mathbf{n}_0$. Experimental data are fitted to Gaussian distribution (black line) to obtain an angular diffusion constant of $D_\varphi = 1.2\ (°)^2 s^{-1} = 3.8 \times 10^{-4}\ rad^2 s^{-1}$. **j**, Translational diffusion of a platelet in a sample plane containing $\mathbf{n}_0$. Gaussian fitting yields the diffusion

constants measured for the translations normal to and along $\mathbf{n}_0$: $D_\perp=0.88\times10^{-3}$ $\mu m^2 s^{-1}$, $D_\parallel=1.5\times10^{-3}$ $\mu m^2 s^{-1}$.